\documentclass[10pt,conference]{IEEEtran}

\usepackage{amsthm}
\usepackage{xspace}
\usepackage{ifthen}
\usepackage{amssymb}
\usepackage{balance} 
\usepackage{listings}
\usepackage[normalem]{ulem} 
\usepackage{amsmath}
\usepackage{array, booktabs}
\usepackage[pdftex,colorlinks=true,pdfstartview=FitV,linkcolor=black,citecolor=black,urlcolor=black]{hyperref}
\usepackage{ifthen}
\usepackage[most]{tcolorbox}
\usepackage{listings}
\usepackage[caption=false]{subfig}
\usepackage{paralist} 
\usepackage{fontawesome}      
\usepackage{adjustbox} 		
\usepackage{pifont} 		
\usepackage{algorithm}
\usepackage{algpseudocode}
\algtext*{EndIf}
\algtext*{EndFor}

\newcolumntype{R}[2]{
    >{\adjustbox{angle=#1,lap=\width-(#2),margin*=0.4em 0em 0em 0em}\bgroup}
    l
    <{\egroup}
}

\let\oldFootnote\footnote
\newcommand\nextToken\relax
\renewcommand\footnote[1]{%
    \oldFootnote{#1}\futurelet\nextToken\isFootnote}
\newcommand\isFootnote{%
    \ifx\footnote\nextToken\textsuperscript{,}\fi}

\newcommand{\id}[1]{$-$Id: scgPaper.tex 32478 2010-04-29 09:11:32Z oscar $-$}

\newcommand{\ie}{\emph{i.e.},\xspace}
\newcommand{\eg}{\emph{e.g.},\xspace}
\newcommand{\etal}{\emph{et al.}\xspace}
\newcommand{\etc}{\emph{etc.}\xspace}
\newcommand{\smell}[1]{\textbf{#1}.}
\newcommand{\Issue}{\emph{Issue:}\xspace}
\newcommand{\Consequences}{\emph{Consequently},\xspace}

\newcommand{\Symptoms}{\emph{Symptom:}\xspace}
\newcommand{\Mitigation}{\emph{Mitigation:}\xspace}

\newcommand{\code}[1]{\texttt{#1}}

\newcolumntype{L}[1]{>{\raggedright\arraybackslash}m{#1}} 

\def\BibTeX{{\rm B\kern-.05em{\sc i\kern-.025em b}\kern-.08em
	T\kern-.1667em\lower.7ex\hbox{E}\kern-.125emX}}
	
\newboolean{showedits}
\setboolean{showedits}{true} 
\ifthenelse{\boolean{showedits}}
{
	\newcommand{\del}[1]{\textcolor{red}{\sout{#1}}} 
	\newcommand{\nbe}[3]{
		{\colorbox{#3}{\bfseries\sffamily\scriptsize\textcolor{white}{#1}}}
		{\textcolor{#3}{\sf\small$\blacktriangleright$\textit{#2}$\blacktriangleleft$}}}
}{
	\newcommand{\del}[1]{} 
	
	\newcommand{\nbe}[3]{}
}

\ifthenelse{\boolean{showedits}}
{
	\newtcolorbox{inserted}{
		title=Inserted text:,
		colframe=blue,colback=blue!5!white,
		breakable,
		leftrule=0mm, 
		bottomrule=0mm,
		rightrule=0mm,
		toprule=0mm,
		arc=0mm, outer arc=0mm,
		oversize
	}
	\newtcolorbox{deleted}{
		title=Deleted text:,
		colframe=red,colback=red!5!white,
		breakable,
		leftrule=0mm, 
		bottomrule=0mm,
		rightrule=0mm,
		toprule=0mm,
		arc=0mm, outer arc=0mm,
		oversize
	}
	\newtcolorbox{refactored}{
		title=Rewritten text:,
		colframe=blue,colback=red!5!white,
		breakable,
		leftrule=0mm, 
		bottomrule=0mm,
		rightrule=0mm,
		toprule=0mm,
		arc=0mm, outer arc=0mm,
		oversize
	}
}{

}

\newboolean{showcomments}
\setboolean{showcomments}{true}
\ifthenelse{\boolean{showcomments}}
{\newcommand{\nbc}[3]{
		{\colorbox{#3}{\bfseries\sffamily\scriptsize\textcolor{white}{#1}}}
		{\textcolor{#3}{\sf\small$\blacktriangleright$\textit{#2}$\blacktriangleleft$}}}
	}
{\newcommand{\nbc}[3]{}
	}

\definecolor{source}{gray}{0.9}
\newcommand{\newevenside}{
	\ifthenelse{\isodd{\thepage}}{\newpage}{
		\newpage
		\phantom{placeholder} 
		\thispagestyle{empty} 
		\newpage
	}
}

\DeclareGraphicsExtensions{.pdf,.png}
\graphicspath{{images/}}

\theoremstyle{definition}

\newcommand{\defref}[1]{\hyperref[{def:#1}]{Definition~\ref*{def:#1}}}

\newboolean{isblinded}
\setboolean{isblinded}{false}
\ifthenelse{\boolean{isblinded}}
{\newcommand\blind[1]{BLINDED\xspace}}
{\newcommand\blind[1]{#1\xspace}}

\begin{document}

\hyphenation{op-tical net-works semi-conduc-tor}

\title{Web APIs in Android through the Lens of Security}

\author{
\IEEEauthorblockN{Pascal Gadient, Mohammad Ghafari, Marc-Andrea Tarnutzer, Oscar Nierstrasz}\\
\IEEEauthorblockA{Software Composition Group, University of Bern\\Bern, Switzerland\\
}}

\IEEEoverridecommandlockouts
\IEEEpubid{\makebox[\columnwidth]{\textbf{Preprint -- SANER 2020}\hfill} \hspace{\columnsep}\makebox[\columnwidth]{ }}

\makeatletter                  
\def\mdseries@tt{m}      
\makeatother                   

\maketitle

\begin{abstract}
Web communication has become an indispensable characteristic of mobile apps.
However, it is not clear what data the apps transmit, to whom, and what consequences such transmissions have.

We analyzed the web communications found in mobile apps from the perspective of security.
We first manually studied 160 Android apps to identify the commonly-used communication libraries, and to understand how they are used in these apps.
We then developed a tool to statically identify web API URLs used in the apps, and restore the JSON data schemas including the type and value of each parameter.

We extracted 9\,714 distinct web API URLs that were used in 3\,376 apps.
We found that developers often use the \code{java.net} package for network communication, however, third-party libraries like \code{OkHttp} are also used in many apps.
We discovered that insecure HTTP connections are seven times more prevalent in closed-source than in open-source apps, and that embedded SQL and JavaScript code is used in web communication in more than 500 different apps.
This finding is devastating; it leaves billions of users and API service providers vulnerable to attack.
\end{abstract}


\section{Introduction}
\label{sec:introduction}

Mobile applications (apps) increasingly rely on web communication to provide their services.
Apps access the internet through web APIs in order to use an increasing number of public web services, or to communicate with private backends.
Researchers have recently studied the use of such APIs in mobile apps, and, for instance, found that  
a large number of web requests are not directly traceable to source code~\cite{Rapoport:2017},
cloud and mail service credentials are hard-coded in the apps~\cite{Zhou:2015},
many web requests are harmful~\cite{Zuo:2017},
many web links targeting well-known advertisement networks impose serious risks on users~\cite{Rastogi:2016},
and lax input validation in many web APIs could compromise the security and privacy of millions of users~\cite{Mendoza:2018}.

We could not, however, find any publicly available tool that researchers can use to study web APIs.
Also, There are several third-party libraries to implement network communication, but existing studies are mainly limited to \code{java.net} APIs.
Finally, 
dissecting the distribution of elements that comprise the web API URLs is never studied, which is necessary for collecting security-related information stored in query keys and values, as well as to fuzz web APIs.

\newpage
We manually studied the use of common web communication frameworks in 160 randomly selected Android mobile apps, \ie more than 4.7\% of the whole dataset, and developed a static analysis tool to investigate whether network communications in 3\,376 closed-source and open-source apps differ. 
We manually inspected the tool's output for 100 random apps, and used the reported URLs to connect to the servers and to investigate their response.
We found eight security code smells, \ie \emph{symptoms in the code that signal the prospect of a security vulnerability}~\cite{Ghafari:2017}, on both ends, dominated by the use of embedded computer languages.
We handcrafted regular expressions to automatically identify the use of those languages, and other languages prevalent on GitHub.

In this work we address the following research questions:

\newcommand{\rqthree}{\emph{What security smells are present in web communication?}\xspace}
\newcommand{\rqfour}{\emph{Which API frameworks are used in Android mobile apps, and what is the nature of the data that apps transmit through these frameworks?}\xspace}

\textbf{RQ$_{1}$}: \rqfour
We identified six different web API communication libraries, and learned that open-source apps rely on simpler request paths including only one or two path segments, while closed-source apps mostly include two or three path segments.
Unexpectedly, the opposite is true for key-value pairs:
Open-source apps frequently use one to three pairs, while closed-source apps mainly use one pair.
Fragments have only been used very sparsely in both types of apps.
We found that open-source and closed-source apps are similar in the choice of web communication libraries, but advertising services are more prevalent in closed-source apps.

\textbf{RQ$_{2}$}: \rqthree
We found eight security smells in the apps and the server software. For instance, 
500 apps use embedded computer languages (\eg SQL, and JavaScript commands) in web API communications, thus introducing the threat of code injection attacks.
A horrific 67\% of the closed-source and 9.5\% of the open-source apps communicate with servers over insecure HTTP connections.
Many apps neglect to use the HTTP strict transport security policy.
Finally, we observed a lack of authentication and authorization mechanisms for services that are supposed to be private.

In summary, this work attempts to shed more light on the use of web APIs in mobile apps, by studying what data the apps transmit, to whom, and for what purpose.
The tool and the obtained results in this study are available online.\footnote{\url{https://github.com/pgadient/jandrolyzer}}

\newpage
The remainder of this paper is organized as follows.
We describe the methodology of our web API mining approach in~\autoref{sec:methodology},
and we present the results of our empirical study in~\autoref{sec:web-api-use-mobile-apps}.
We report numerous web API security smells in~\autoref{sec:web-api-security-smells}.
Finally, we recap the threats to validity in~\autoref{sec:threats-to-validity}, and we summarize related work in~\autoref{sec:related-work}. We conclude this paper in~\autoref{sec:conclusion}.

\section{Web API Mining}
\label{sec:methodology}

We manually inspected Android apps to identify what APIs developers use to call web services, and how they are used.
Then we took advantage of this information to develop a tool to automatically extract the web API URLs and their corresponding HTTP request headers statically from the apps.

\subsection{Library Inspection}
An Android app can call a web API either with the help of the built-in Java classes, or by using external third-party libraries.
We consulted the official Java and Android documentations to compile a list of built-in APIs that are relevant to network communication, and to establish how these APIs are used.
We mainly focused on the \code{java.net} package, which includes  a number of classes such as \code{Socket}, \code{HttpsURLConnection}, and \code{URLConnection} to implement network-related operations

Next, we manually inspected 160 randomly selected apps from a dataset of 3\,376 apps (see \autoref{sec:web-api-use-mobile-apps}) that request Android's \code{INTERNET} permission to investigate what third-party libraries they may use for web communication, and how.
These libraries are often built on top of the built-in Java network APIs.
Therefore, we first checked whether a call to such Java APIs exists, and, if so, we checked whether the call belongs to the app or an external library.
For each library, we studied the documentation, and investigated how developers use the library in each app, \eg to construct URLs, and to attach headers to web requests.
During the inspection of each app, we collected the web API URLs and any data that are transmitted to the servers to determine if what we collect from the source code is actually helpful to issue valid requests.

In this study, besides the native Java network libraries, we found that libraries such as \emph{Apache HttpClient}, \emph{Glide}, \emph{Ion}, \emph{OkHttp}, \emph{Retrofit}, and \emph{Volley} are used in the apps.

While studying the use of web communication libraries, we also noticed that besides the built-in \emph{org.json} package, developers often use two external libraries, namely \emph{Gson} and \emph{Moshi}, for parsing and manipulating JSON (JavaScript Object notation) data, which is commonly used for data exchange in web services.


\subsection{API Miner}

We then developed a tool that leverages our finding in the library inspection phase, and statically analyzes apps to extract web API URLs, query keys and the corresponding values where applicable.
The tool takes the following steps:

\subsubsection{Decompilation}

Given an APK file, the tool first decompiles the app using the command line version of the \emph{JADX} decompilation tool.\footnote{\url{https://github.com/skylot/jadx}}
A successful decompilation will provide us with a project folder that contains decompiled Java source code of the app and the resource files.
Although decompilation errors are common, JADX is quite robust and produces code with a correct syntax.
In particular, method declarations
and class structures remain intact with comments in place where the decompilation did not succeed completely.

The tool uses the \emph{JavaParser} framework to create an abstract syntax tree (AST) for every \code{.java} file within the project.\footnote{\url{https://javaparser.org}}
When the actual source code of an  app is available, we use the information from the build and configuration files to accurately inject specific library versions into the \emph{JavaParser} framework to enable the resolution of library dependencies in the subsequent app analysis.
If the desired library version is unavailable in our collection, the next available more recent version is added instead.
Closed-source apps (\ie APKs) do not require those dependency injections as they already contain the required code themselves.

\subsubsection{Detection and Extraction}

In principal, we need to track flows of data in relevant APIs, and several static analysis frameworks exist to track data flows in Android apps.
Nevertheless, in our experience as well as according to recent studies, these tools may not perform as described in the relevant papers~\cite{Qiu:2018, Pauck:2018, Corr18a}.
We therefore decided to implement our own lightweight analysis tailored to reconstruct web APIs in the code.

The tool traverses the AST to identify APIs, \ie \code{MethodCallExpression} nodes, that are used to access web APIs in a network library.
For each method call, it recursively resolves the nodes on which the API depends, \eg the object on which the method is called, and its parameters.
In detail, we rely on the \emph{JavaSymbolSolver} framework to associate a variable in the code to its declaration.\footnote{\url{https://github.com/javaparser/javasymbolsolver}}
We track all \code{Assignment}, and \code{MethodInvocation} constructs on each variable in each relevant \code{VariableDeclaration} node.
Moreover, depending on the target library, the tool also tracks implicit dependencies, \eg the annotation-driven dependency injection.

URL and header\footnote{The HTTP request header is a plain text record providing input details for the web API request.} construction largely depend on string concatenation. We therefore support the extraction of strings that are built using the \code{StringBuilder.append()} method, the \code{String.concat()} method, and the ``\code{+}'' operator.

\subsubsection{Reconstruction}
All web API URLs and JSON data structures that contain at least one unresolved value are further processed in the reconstruction stage.
We set the value of variables whose types are number or boolean to \code{0} and \code{true}, respectively.
For those variables (\ie JSON or query keys) whose types are \code{String}, and for which we did not find a concrete value during the extraction, we compute the \emph{Jaro-Winkler} similarity distance~\cite{Winkler:1991} between the variable names and every variable declaration in the code.
In the end, for each successful analysis, the tool reports the web API, as shown in \autoref{lst:output-url-example}, and the corresponding request headers, as shown in \autoref{lst:json-output-example}.

\begin{lstlisting}[label=lst:output-url-example,caption=The tool's output for a successful web API extraction,xleftmargin=10pt]
Path: 
/Users/marc/...
Library: 
com.squareup.retrofit
Scheme: 
http://
Authority: 
retrofiturl.com
Base URL: 
http://retrofiturl.com
Endpoints: 
    Path: api/loadUsers
    Queries: 
        Query key: position, query value: <String>
        Query key: order, query value: <String>
    Fragments: 
    HTTP Methods: 
        HTTP Method: GET
\end{lstlisting}

\begin{lstlisting}[label=lst:json-output-example,caption=The tool's output for a successful JSON object extraction,mathescape=true,xleftmargin=10pt]
Path: 
.../User.java
Library: 
com.squareup.moshi
JSON Object: 
{"address":{"street":"<STRING>",
	"number": <NUMBER_INT>"},"name":"Bob"}
\end{lstlisting}

\subsubsection{Evaluation}
We performed a lightweight evaluation of the tool on 10 open-source and 10 closed-source apps randomly selected from our dataset.
In each app, we manually searched for the terms ``http://'' and ``https://" in the (decompiled) source code.
For each finding, we evaluated which entries were related to web APIs, and then tried to understand what are the URLs and the other request parameters.

We manually identified 24 distinct URLs for web APIs in the apps, of which 21 were found in the Java source code.
The tool reported 39, of which 18 URLs referred to web services:
17 were amongst the URLs identified manually, and the tool uncovered one new case that was overlooked due to complex string concatenation.
The tool achieved a precision of 46\% and a recall of 80\%.

There are several reasons for the tool missing the remaining seven URLs, such as URLs in open-source apps being hidden in build scripts and XML resource files rather than Java code, and incomplete library injections for closed-source apps.

The tool reported 21 URLs that did not refer to a web service. 
In particular, 18 URLs referred to static HTML pages, and three suffered from invalid reconstruction.

\subsection{Security Checks}
We inspected the result of the tool on a random set of 100 apps in order to identify security smells in the code relevant to web API communications.

We implemented lightweight detection strategies for these smells, mainly using regular expressions.
For instance, using search terms such as username, password, \etc we could find hard-coded passwords, tokens, and insufficiently protected authorization schemes in the results.

\begin{lstlisting}[label=lst:computer-languages,caption=Examples of embedded computer code in app source strings,mathescape=true,xleftmargin=10pt]
HTML:
String uiElement = "<html><body>" + 
    $\hookrightarrow$ jsonObj.getText() + "</body></html>";

JavaScript:
String customScript = jsonObj.getResponse();

SQL:
String queryParameter = "SELECT * FROM weather";
\end{lstlisting}

In many apps we found code from various computer languages embedded in Java strings, such as that shown in~\autoref{lst:computer-languages}, thus potentially exposing the app or the server to code injection attacks.
We compiled a list of commonly used computer languages based on our own findings, and the scripting languages found in the top ten used programming languages on GitHub.\footnote{\url{https://github.com/oprogramador/github-languages}}
For each language, we pragmatically developed regular expressions inspired by the relevant language specifications, with the aim to match as many occurences as possible.
With these regexes, shown in~\autoref{tab:list-of-regex-strings}, we counted the key identifiers for each language in each app report, to detect usages of embedded languages in the web communications.

\begin{table}[!htbp]
\centering
\caption{Regular expressions used to detect computer languages}
\scriptsize
\begin{tabular}{ p{1cm}  p{2.8cm} | p{1cm}  p{2.3cm} }
\textbf{Language} & \textbf{Regular expressions} & \textbf{Language} & \textbf{Regular expressions}\\ \hline
Bash 	& sh[ ]+ 	& SQL 	& alter[ ]+table \\
 	& \%.sh 	&  	& create[ ]+.*index \\ \cline{1-2}
HTML 	& \%\textless[ ]*html[ ]*\%\textgreater 	&  	& create[ ]+.*table \\ \cline{1-2}
JavaScript 	& function[\^{}\%(]*\%([\^{}\%)]*\%) 	&  	& create[ ]+.*trigger \\
 	& \%\textless[ ]*script 	&  	& create[ ]+.*view \\ 
 	& js[ ]*= 	&  	& delete[ ]+from \\ \cline{1-2}
PHP 	& \%\textless\%? 	&  	& drop[ ]+index \\ \cline{1-2}
Python 	& import[ ]+\%(.*\%) 	&  	& drop[ ]+table \\ \cline{1-2}
Ruby 	& require[ ]*\%(.*\%) 	&  	& drop[ ]+trigger \\
	&	&  	& drop[ ]+view \\ 
	&	&  	& insert[ ]+.*into \\ 
	&	&  	& replace[ ]+into \\ 
	&	&  	& select[ ]+.*[ ]+from \\
	&	&  	& update[ ]+.+[ ]+set \\
\end{tabular}
\normalsize
\label{tab:list-of-regex-strings}
\end{table}

In a subsequent step, we issued requests to each of the URLs extracted from the entire dataset, and observed unexpected responses, \eg stack traces, error messages, or status information, disclosing sensitive information regarding the API implementation, running software, or server configuration.

\section{Study Result}
\label{sec:web-api-use-mobile-apps}
We investigated the use of network communication in Android mobile apps.
In particular, the focus is on the use of libraries, and the request characteristics.

We randomly collected apps that use internet.
For closed-source apps we mined the free apps on the \emph{Google Play} store, and for the open-source apps we relied on the \emph{F-Droid} software repository.\footnote{\url{https://f-droid.org}}
For each app, we removed the duplicates, \ie apps with the same package identifier, but different version numbers, and kept only the most recent version of the app.
In the end, we collected 17\,079 closed-source, and 432 open-source apps.

We applied our tool to these apps, and restricted each app analysis to 30 minutes processing time, with a node resolution limit of 15 iterations on a machine with two AMD Opteron 6\,272 16-core processors and 128 GB of ECC memory.
The tool could completely analyze 293 open-source apps, and 2\,410 closed-source apps.
We also included the partial results of the apps whose analyses were incomplete, resulting in a total analysis result of 303 open-source, and 3\,073 closed-source apps in our dataset.
Only 2\,587 apps (15\%) were successfully decompiled, due to crashes of the tool caused by various bugs, and incomplete feature support, \eg reflection, native code, and customized app configurations.

The apps in our dataset come from 48 different Google Play store categories.
Most of them belong to \code{EDUCATION} (317 apps) and \code{TOOLS} (292 apps), however, a majority (574) have a \code{GAMES}-related tag.
Interestingly, work-related apps are common in our dataset (335 apps).
The top five categories whose apps contain the largest number of distinct web API URLs are \code{EDUCATION} (1\,555 URLs), \code{LIFESTYLE} (1\,027 URLs), \code{BUSINESS} (995 URLs), \code{ENTERTAINMENT} (704 URLs), and \code{PRODUCTIVITY} (619 URLs).

We present our findings in the following, and conclude each focal point with a short discussion, which entails similarities or differences in open-source and closed-source apps.

\subsection{Communication Libraries}

We investigated the distribution of the seven different communication libraries in 3\,376 apps in our dataset.

\subsubsection{Result}

In \emph{open-source apps}, we found that each app uses up to four network libraries.
The \code{URLConnection} (37\%), \code{Http\-URL\-Con\-nect\-ion} (24\%), \code{Socket} (9.1\%), and \code{HttpsURLConnection} (6.0\%) classes included in \emph{java.net} are the preferred choice of open-source developers, especially \code{URLConnection} and \code{Http\-URL\-Con\-nect\-ion} are omnipresent in projects.
When considering third party network libraries, we found that \emph{OkHttp} and \emph{Retrofit} (each 5.6\%) are used the most.
It is interesting to see that libraries with specific support for image downloads are similarly used, \ie \emph{Glide} and \emph{Volley}.
The \emph{Ion} library is used only in three apps (1.0\%).

In \emph{closed-source apps} each app uses up to seven network libraries.
We found that the classes included in \emph{java.net} such as \code{URLConnection} (42\%), \code{HttpURLConnection} (34\%), \code{Socket} (10\%), and \code{HttpsURLConnection} (4.3\%) are the preferred choice.
Interestingly, the \emph{OkHttp} library is the most commonly used third-party library even surpassing the well-known \emph{Glide} and \emph{Retrofit} libraries.
We found \code{org.apache.httpcomponents} and \code{com.loopj.android} are the two least used network libraries contributing only 0.9\% and 0.5\%, respectively.

\subsubsection{Discussion}
We realized that one to three classes are usually responsible for network communication in an app.
In open-source apps we found the use of up to four network libraries in each app, and in closed-source apps it was up to seven.
Although each library may provide specific features, \eg JSON parsing, HTTP connection management, image caching, \etc, we expect the reason for the use of multiple libraries in an app is that many developers use the code snippets from other projects or online information sources.

We found fewer \emph{java.net} libraries in open-source apps compared to closed-source apps.
During decompilation, the bundled libraries are decompiled together with the app code.
Therefore, what the tool reports is not only the network calls in the app code, but also the network APIs on top of which the third-party libraries are developed.
However, this is not the case for the open-source apps whose dependencies are defined in Gradle, and are dynamically injected without adding the actual code to the project itself.

The libraries \emph{Ion} and \emph{Volley} have been used only in open-source apps, while \emph{HttpComponents} and \emph{LoopJ} have been used only in closed-source apps.
Surprisingly, we did not find any instances of the well-known \code{AndroidHttpClient} and \code{SSLSocket} classes.
Finally, the use of \emph{Glide}, which supports exhaustive image downloading and caching features, seems much more prevalent on closed-source apps.

\subsection{The Nature of Web API Requests}
\label{subsubsec:use-of-urls-and-json-schemes}

Based on the analysis results for the apps in our database, we investigated the structure, dissemination and use of 13\,276 web API URLs, of which 9\,714 were unique.

\subsubsection{Open-source Apps}
The tool extracted 1\,533 URLs from the open-source projects.
We found that the majority of web APIs consist of one or two queries or path segments.
We only found up to one fragment per web API.
We further found that 209 web APIs exist with paths consisting of four or five segments to distinguish between resources (the average number of segments in the web APIs is 2.36).
Nevertheless, web APIs using more than five elements are rare.
Web APIs contain an average of 2.3 key-value pairs in queries.
The data do not follow a normal distribution.

Surprisingly, the top base URL was \url{https://github.com}, which we observed 29 times (1.8\%).
Likewise, \emph{Google} services have been widely used, \eg \url{https://play.google.com} or \url{https://plus.google.com}, of which the tool could spot 42 instances (2.7\%).
Rather at the end of the ten most commonly used base URLs the tool found the \emph{OpenWeatherMap} API \url{http://openweathermap.org} (7, 0.4\%) and the \emph{Twitter} social network API \url{https://twitter.com} (6, 0.3\%).

Furthermore, we found that the \code{https} URL scheme (1\,012 occurrences, 66\%) is much more commonly used than its insecure counterpart \code{http} (521 occurrences, 33\%).

\subsubsection{Closed-source Apps}
The tool extracted 11\,743 URLs from closed-source apps.
We found that the majority of web APIs consist of one or two queries or path segments.
On a second look, we observed that web APIs with two path segments are most prevalent.
We further discovered that 2\,116 web APIs exist with paths consisting of four to eight path segments to distinguish between resources (the average number of segments in the web APIs is 2.44).
Nevertheless, web APIs using more than four elements are rare.
Additionally, we could identify that URL fragments are seldomly used in web APIs; although we found up to seven fragments in a single web API URL, we only discovered 183 web APIs in total using this feature, \ie 1.5\%.
Web APIs, on average, contain 2.9 key-value pairs in queries.
The data do not follow a normal distribution.

Interestingly, all the most common URLs we could retrieve were pointing towards \emph{Google} services.
The top URL, \url{http://schemas.android.com}, was observed 1\,303 times (11\%).
Two of the observed URLs were related to advertising distribution services, \ie \url{http://media.admob.com} (283, 2.4\%) and \url{https://pagead2.googlesyndication.com} (271, 2.3\%).\footnote{\emph{Google AdMob} is a popular advertising platform that provides SDKs to developers to integrate \emph{Google} ads into their own apps to increase revenue.}

We found that the \code{http} URL scheme (7\,208 occurrences, 61\%) is much more prevalent than its secure counterpart \code{https} (4\,531 occurrences, 38\%).
Besides findings of the two common schemes we found few appearances of the \code{ws} (WebSocket) protocol (4 occurrences, 0.0\%), which provides (unprotected) full-duplex communication on top of HTTP TCP connections.

\subsubsection{Discussion}
The number of used path segments and query keys are an indicator for the complexity of a specific request.
Servers usually reject requests with incomplete or flawed parameter configurations, and thus the task of sending a successful request becomes harder the more path segments and query keys are involved.

Open-source apps relied on simpler request paths including only one or two path segments, while closed-source apps mostly included two or three path segments.
Unexpectedly, the opposite is true for key-value pairs:
Open-source apps frequently use one to three pairs, while closed-source apps majorly use one pair.
Fragments have only been used very sparsely in both types of apps.

We did not expect to observe a difference between open-source and closed-source apps.
Moreover, we did not expect to find many complex requests, because the idea of providing APIs is that they can be used by other developers who presumably prefer an easy to use interface.
We conclude that the majority of the APIs provide a simple interface and are rather straightforward to access.

While the open-source apps contained no advertising services in the ten most used base URLs, the closed-source apps heavily used such services.
We expect that the ``Freemium'' price model, \ie installation of apps is free but the user must later watch ads or pay a fee, is a major enabler of this setting.

The open-source community prefers the \emph{Twitter} social network over \emph{Facebook}.

We found one major difference in the URL schemes used in the apps.
Open-source apps principally rely on secure \code{https} connections (66\%).
In contrast, closed-source apps largely use the insecure \code{http} protocol (60\%).
We see here much potential for improvement through stricter rejection of apps using insecure connections.
The more efficient, but more complex WebSocket protocol seems to be out of interest for the majority of developers.

\subsection{Security Risks}

We studied the kinds of data communicated through web APIs, and found that both credentials (\ie user name and password combinations) and embedded code were very common in the web communications.
As the former has been reported on extensively in the past, we focus here on the latter.

\subsubsection{Open-source Apps}

The tool extracted 458 JSON schemes in which
\code{STRING} is the most used value type with 1\,197 occurrences, followed by \code{NUMBER} with 234 occurrences.

We found that SQL (91\%, 10 affected apps) is by far the most used embedded language.
HTML (5.5\%, 2 affected apps) and JavaScript (2.7\%, 1 affected apps) are very rare.
No instances of other embedded languages were detected.

\subsubsection{Closed-source Apps}

The tool extracted 14\,606 JSON schemes where \code{STRING} is the most used value type with 40\,017 occurrences, followed by \code{BOOLEAN} with 5\,640 occurrences.
\code{NUMBER} and \code{NULL} only represent a minority with 2\,389 and 1\,483 occurrences, respectively.

In contrast to open-source apps, we observed that JavaScript (76\%, 170 affected apps) is very prevalent, and SQL (23\%, 476 affected apps) is used less, but still frequently.
HTML code is almost non-existent (0.7\%, 27 affected apps).

\subsubsection{Discussion}

We found that the use of tokens in open-source apps is not as common as in closed-source apps.
One explanation could be that the fees associated to web services do not pay off for open-source apps which mostly do not generate any revenue.

Several embedded languages are actively used within mobile apps.
While SQL is relatively common in both open-source and closed-source apps, JavaScript is much more commonly used in the latter.

\section{Web API Security Smells}
\label{sec:web-api-security-smells}

In this section, we present the security smells that we found in web communication during investigation of the tool's results, by manually investigating 100 apps, and by analyzing the responses from requests to each of the 9\,714 web API URLs extracted from apps in our dataset.
We classify the smells into client side (\ie within mobile apps), and server side (\ie on the API servers).
For each smell we report the security \emph{issue} at stake, the potential \emph{consequences} for users, the \emph{symptom} in the code (\ie the code smell), and the recommended \emph{mitigation} strategy of the issue, principally for developers.

We used the results from the manual analysis explicitly to identify security issues, but not to perform any quantitative evaluation.
In this section, we do not report any number of occurrences found in the tool's results, because those either have been discussed in the previous section, or the task would require additional research to gather qualitative results.

In our analysis, we could identify eight web API security smells, of which three were in apps and five in server implementations.
Two of the three web API app security smells could be mitigated, if only secure HTTPS channels would be used for communication.
We have not yet reported our findings to developers or marketplaces.

\subsection{Client-side}
\label{subsec:apps-list-of-smells}

We identified three client-side web API security code smells.
\begin{itemize}
\item \smell{Credential leak}
We found hard-coded API keys, login information, and other sensitive data, \eg email addresses, in the source code.
Several of the retrieved data were valid at the time of our investigation:
We could access \emph{Google Maps}, \emph{Mapquest}, \emph{OpenWeatherMap}, the \emph{San Francisco transit} API, and a \emph{Telegram} bot.
\\
\Issue
Credentials issued to app vendors are prevalent in apps that use web APIs, and they are statically stored in the Java software to perform the queries.
However, the software can be decompiled into source code, which renders the data extraction trivial. 
\\
\Consequences
web services can be misused by people who have gained access to unique credentials.
Such services allow impersonation, phishing, information leaks, fake messages, or financial infringements for the app developers due to API overuse or lockdowns.
\\
\Symptoms 
Query keys like \code{key}, \code{token}, \code{user}, \code{username}, \code{password}, \code{pw} are used in web requests and the corresponding values are statically stored in the apps.
\\
\Mitigation 
Developers should avoid using access tokens and logins of corporate accounts for apps.
Instead, a unique child token based on the corporate token should be assigned to every user.
If this option is unavailable, web relay APIs can be provided to the apps which forward the requests to the final destination without disclosing any credentials.

\item \smell{Embedded languages}
We found apps that assemble CSS, HTML, or JavaScript code programmatically using external input.
In many apps, such constructed code is executed within a \code{WebView} or Android's UI framework, which is inspired by \emph{Java Swing} and supports HTML elements.
Similarly, we found assembled SQL statements that are executed in the local SQLite database engine.
In two apps we found assembled shell commands sent over an SSH connection.
\\
\Issue 
An attacker could gain control over the app's visual representation, the behavior, the data storage, or the corresponding server by exploiting such code.
Shell commands such as \code{String command = "touch /home/" + username + "/.toolConfig/configuration";} allow an adversary to execute commands on a server by letting the variable \code{username} be \code{;echo 'executes on server';}.
\\
\Consequences 
for HTML and CSS, an attacker could change the appearance of existing web elements to make space for additional ones, \eg by reducing the font size of existing text to make it impossible to read and at the same time injecting additional text in regular size.
Such changes can trick users into taking unintended actions.
With JavaScript, an attacker could gain access to the \emph{Document Object Model (DOM)} of the app's webpage and extract or alter the visible content.
Such changes expose sensitive user data, or mislead users through altered information.
SQL allows adversaries to perform arbitrary actions on the database, \eg altering and deleting existing data, or inserting new data.
This leads to data loss, corruption, or leaks for the users.
Through shell commands an adversary could potentially gain elaborated remote access to the server's operating system.
Threats range from DoS attacks to sensitive user information leaks and corporate network infiltrations by disabling security measures and installing malicious software on the server.
\\
\Symptoms 
At least one statement is manually assembled with the help of external data, \eg \code{"<html><body>" + example + "</html></body>"} or \code{"color:" + color + ";"}.
HTML/CSS: common tags or properties occur, \eg \code{"<html>"}, \code{"<body>"}, or \code{"color:"}.
JavaScript: identifiers exist in the app, \eg \code{function()}, \code{<script}, \code{js=}.
SQL: keywords are used in the app, \eg \code{SELECT}, \code{INSERT}, \code{UPDATE}, \code{DELETE}, \code{REPLACE}, \code{TRUNCATE}.
Shell: commands are not trivial to detect, because developers use a variety of different commands, \eg \code{sudo}, \code{rm}, \code{cp}, \code{mv}, \code{ls}, \code{exec}, \code{attrib}, \code{chmod}, \code{touch}, \etc
\\
\Mitigation 
Developers should not use external input when assembling embedded languages, but try to embed the content into the app installation or update package.
Static code should be used whenever possible.
If dynamic code is required, the built-in sanitizing classes must be used, \eg \code{PreparedStatement} for SQL code.
User input should \emph{never} be trusted.
In general, \emph{any} untrustworthy input must not be used before it is properly escaped and sanitized.

\item \smell{Insecure transport channel}
Web API communication relies on HTTP or HTTPS; both variants exist in apps.
\\
\Issue 
HTTP does not provide any security; neither the address, nor the header information or the payload are encrypted.
\\
\Consequences 
any attacker with access to the transmitted data can read or alter all plain text messages.
User data leaks, corruptions, losses, or impersonation are probable.
\\
\Symptoms 
HTTP URLs are used to establish connections to web APIs.
\\
\Mitigation 
HTTPS instead of HTTP URLs must be used for any web communication.

\end{itemize}

\subsection{Server-side}

For every collected API in our dataset,
we accessed the corresponding web server and stored the response.
We were particularly interested in information such as operating system identifiers, used software modules, and version numbers, which we could initially identify during the manual analysis of a sample of the server responses.
We then crafted a number of search queries
to detect occurrences of such features and applied them to our dataset.

We have identified five server-side web API security code smells.

\begin{itemize}
\item \smell{Disclosure of API implementation code}
Error messages provide valuable information regarding the implementation of a running system.
We found web APIs that leak internal error states and use \emph{status codes} in a different way than what is specified by the RFC7231.\footnote{\url{https://tools.ietf.org/html/rfc7231}}\footnote{Although HTTP servers should reply with the status code \code{200} to indicate a successful request, we noticed that some servers use this status code when an error has occurred.}
\\
\Issue 
Error messages that include the relevant stack trace are transmitted as plain text in the server's message response body.
Such a message reveals information like the used method names, line numbers, and file paths disclosing the internal file system structure and configuration of the server.
\\
\Consequences 
adversaries can obtain detailed information about the service implementation, which may lead to an exploit.
\\
\Symptoms 
When an invalid request is received, a server responds with a detailed error message containing information that is not required by any user of the API.
\\
\Mitigation 
If the used framework provides an option to turn off diagnostic or debug messages: this feature should be used.
Otherwise, an API gateway in between the client and the server should filter such responses and deliver regular HTTP \code{500} messages to the client instead.

\item \smell{Disclosure of version information}
Besides useful connection parameters, HTTP headers provide information regarding the software architecture and configuration of a running system.
We spotted in the reported HTTP headers version information of web server daemons and API implementation frameworks.
\\
\Issue 
We encountered outdated software that suffers from severe security vulnerabilities.
For instance, we observed a server that returned \code{X-Powered-By: PHP/5.5.23} in the response header.
This PHP version is at the time of writing more than 6 years old, and a quick search in the Common Vulnerabilities and Exposures (CVE) database showed that this framework suffers from 69 known security vulnerabilities, six of which received the most severe impact score of 10.\footnote{\url{https://www.cvedetails.com/vulnerability-list/vendor_id-74/product_id-128/version_id-183021/PHP-PHP-5.5.23.html}}
\\
\Consequences 
the vulnerabilities range from simple DoS attacks, access control bypassing, and cross-site scripting to arbitrary code execution on the server.
\\
\Symptoms 
One of the following header keys exists in the response header: \code{engine}, \code{server}, \code{x-aspnet-version}, or \code{x-powered-by}.
\\
\Mitigation 
If the used software provides an option to turn off the publishing of version information: this feature should be used.
Otherwise, an API gateway in between the client and the server should remove the affected keys and deliver messages with sanitized HTTP headers to the client instead.

\item \smell{Lack of access control}
Authentication by a user name and a password provides tailored experiences to end users, \eg individual chat logs or friend lists, and at the same time enables access control to separate and protect sensitive user data.
\\
\Issue 
The access to sensitive data or actions is not restricted by a sane authentication mechanism such as a user name and password pair, instead, easy-to-forge identifiers or no identification data at all are used to secure the access.
We found several APIs that did not use any authentication or authorization mechanisms, although they host sensitive data, \eg for car rental services and accounting.
In one app we found code to access an exposed SQL database interface.
\\
\Consequences 
every internet user can access sensitive data or perform unauthorized actions including the reading, modification, and deletion of arbitrary user data.
We could access information from such APIs, \eg real-time location data of rental cars and transaction histories on different bank accounts.
In one case, we were also able to create new users in the system.
Exposing database or other interpreter interfaces with broken authentication allows adversaries to execute arbitrary statements on the server.
\\
\Symptoms 
A web API server hosts sensitive data or provides actions which would require elevated access rights.
The server responds without asking for any login information, that is, no HTTP headers or keys related to personal information are used in the API, \eg \code{username}, \code{password}, or \code{pw}.
The server requires query keys with names of programming languages, \eg \code{sql}, and responds when such variables hold a statement in that language, \eg \code{SELECT table\_name FROM all\_tables;}.
The decision finding of data sensitivity or elevated actions is non-trivial and involves manual reasoning~\cite{Yee:2017}.
Therefore, we cannot infer general purpose terms.
\\
\Mitigation 
Application architects have to implement authentication, favorably multi-factor authentication, whenever sensitive data or elevated operations are involved in the process.
All user data, and location data in general, have to be considered as sensitive.
Developers should never expose interpreter interfaces to a web service without prior authentication and input validation.
REST interfaces for specific tasks should be created, preferably each using static statements that do not rely on any user input.

\item \smell{Missing HTTPS redirects}
In contrast to HTTPS, HTTP does not provide any security:
neither the URL, nor the header information and embedded content are encrypted.
We found servers that do not redirect the clients to encrypted connections although they would have been supported.
\\
\Issue 
Web API servers do not redirect incoming HTTP connections to HTTPS when legacy apps try to connect, or users manually configure a URL without adding a proper \code{https://} prefix.
\\
\Consequences 
the transmitted data remains visible and changeable to anyone within the communication path.
\\
\Symptoms 
For an HTTP web API request, a server does not deliver an HTTP \code{3xx} redirect message which points to the corresponding HTTPS implementation of the web API.
\\
\Mitigation
A server should not offer legacy HTTP services.
If they are still required due to legacy clients with hardcoded HTTP URLs, redirects should be provided to guide all clients to the secure version.

\item \smell{Missing HSTS}
HTTP header information is used to properly set up the connection by specifying various communication parameters, \eg the acceptable languages, the used compression, or the enforcement of HTTPS for future connection attempts, a feature which is called \emph{HTTP Strict Transport Security (HSTS)}.
HSTS provides protection against HTTPS to HTTP downgrading attacks, \ie when a user once accessed a web resource in a secure environment (at home or work), the client knows that the resource needs to be accessed \emph{only} through HTTPS.
If this is not possible, \eg at an airport at which an attacker tries to perform MITM attacks, the client will display a connection error.
Hence, HSTS should be used in combination with HTTP to HTTPS redirects, because the HSTS header is only considered to be valid when sent over HTTPS connections.
We found servers that do not enforce clients to remain on the secure channel for future requests.
\\
\Issue 
Servers do not leverage the HSTS feature.
\\
\Consequences 
in unprotected public networks or networks under external supervision, if an attacker sets up a fake gateway which runs SSLsniff,\footnote{\url{https://github.com/moxie0/sslsniff}} the provided services remain vulnerable, because transmitted data is visible and changeable.
\\
\Symptoms 
A server does not deliver the HTTP HSTS header \code{Strict-Transport-Security: max-age=31536000; includeSubDomains} for an HTTPS request.
\\
\Mitigation 
In combination with HTTP to HTTPS redirects, the HSTS header should be used in all HTTPS connections.

\end{itemize}

\section{Threats to Validity}
\label{sec:threats-to-validity}

The main threat to validity is the completeness of this study, \ie it is not guaranteed that we found all major libraries used for web communication in Android apps.

There may be bias in the apps that we selected for this study.
We included all open-source apps that were available on \emph{F-Droid}, but they may not be representative of the whole open-source app community.
We collected random closed-source apps that were freely available on the \emph{Google Play} store, but paid apps or the apps on third-party stores may have different characteristics.

We only mined web APIs that were available in the source code; our tool suffers from the inherent limitations which come with static source code analysis.
We developed a lightweight analysis, which is not path sensitive.
We opted for this design because, during manual inspection of  network APIs in the apps, we noticed that these APIs are usually free of conditional statements and loops.
Furthermore, we had to decompile closed-source apps for analysis, which introduces further threats to the validity of our results.
For instance, the app code and its library code are not easy to discern automatically, and therefore the libraries in such apps may have influenced our findings.

We did not evaluate how complete the tool results are for every app, but just a small number. 
There is a threat to construct validity through potential bias in our expectancy.
However, we examined the tool results for 50 apps, and confirmed that 90\% led to successful communication with the web APIs.

\section{Related Work}
\label{sec:related-work}

In previous work, we defined the notion of security code smells and investigated their appearance in 46\,000 closed-source Android apps from the official market~\cite{Ghafari:2017}.
We identified 28 different security smells in five different categories, and found that \emph{XSS-like Code Injection}, \emph{Dynamic Code Loading}, and \emph{Custom Scheme Channel} are the most prevalent smells.
In a follow-up work, we studied the prevalence of \emph{Inter-Component Communication (ICC)}-related security smells in more than 700 open-source apps, and manually inspected around 15\% of the apps to assess the extent to which identifying such smells uncovers ICC security vulnerabilities~\cite{Gadient:2018}.
We found that almost all apps suffer from the \emph{Common Task Affinity} smell, and that \emph{Unauthorized Intent} and \emph{Custom Scheme Channel} are prevalent among mobile apps.
Furthermore, we discovered that updates rarely have any impact on ICC security, however, in case they do, they often correspond to new app features. 
The manual investigation of 100 apps showed that our tool successfully found many different ICC security code smells, and about 43\% of them in fact represent vulnerabilities.


Zhou~\etal harvested free email and Amazon AWS cloud service credentials with their tool \emph{CredMiner} from more than 36\,500 apps from various Android markets~\cite{Zhou:2015}.
In their case studies, they mention unprotected credentials within the app's source code, obfuscated credentials using a \emph{Base64} encoding, and encrypted credentials, however, in those cases the decryption key has also been found in the app's source code.
They alarmingly found that more than every second app using such a service leaked the developers' credentials in the apps' source code.
Making matters worse, more than 77\% of those collected credentials were valid at the time of the experiment.
Such credentials will present a massive threat in the mid-term future, as many of those credentials cannot be easily replaced without temporarily abating the experience of millions of users, but in the meantime they can be easily exploited by attackers.


Rapoport~\etal studied web requests in Android apps~\cite{Rapoport:2017}.
They demonstrated that a large number of web requests are not immediately traceable to source code and need dynamic analysis.
For instance, URLs may originate in app resources, \eg XML files or Gradle build scripts, they may stem from the content received from previous web requests, or they might be assembled by JavaScript code at run time.
In contrast, a significant proportion of URLs are only detected by static analysis: the dynamic analysis may simply fail to produce desired results due to a lack of code coverage during instrumentation.


Zuo~\etal analyzed 5\,000 top-ranked apps in Google Play and identified 297\,780 URLs~\cite{Zuo:2017}.
They fed the URLs to a harmful URL detection service at VirusTotal, and found 8\,634 harmful URLs.
The harmful URLs have been classified into three different non-distinct threat categories: phishing (23\%), malicious sites (37\%), and malware (43\%).
For the malware category, one interesting example they mention is an APK file download triggered by an app, which itself tries to obtain superuser access to the device by exploiting Linux kernel vulnerabilities.


Mendoza~\etal studied the inconsistencies in input validation logic between apps and their respective web API services~\cite{Mendoza:2018}.
They developed a tool to extract requests to web API services from an app, and to infer sample input values that violate the implemented constraints found in the app, such as email address or JSON content validation executed on the client side.
They then analyzed app-violating request logic on the server side via black box testing.
From a set of 10\,000 popular Android apps, they found 4\,000 apps that do not properly implement input validation for web API services.
Investigation of web API hijacking vulnerabilities in 1\,000 apps showed that the security and privacy of millions of users are at risk.


In summary, we could not find any publicly available tool that researchers can use to study web APIs.
Also, existing work usually focused on the use of \code{java.net} APIs, and did not study several third-party libraries to implement network communication in Android apps.
Finally, to the best of our knowledge, dissecting the distribution of elements that comprise the web APIs, and the use of embedded languages, is never studied.

\section{Conclusion}
\label{sec:conclusion}

We manually reviewed 160 Android apps to compile a list of commonly used network and data conversion libraries and to learn how they are used in these apps.
Based on our findings, we developed a lightweight static analysis tool that identifies network-related APIs, and extracts communication information such as the web APIs, and the associated JSON headers.
With the help of our tool we successfully analyzed the network-related information within 450 closed-source and open-source apps.
We found that in both open-source and closed-source apps network communication is mainly developed using \emph{java.net} classes.
Amongst the third-party libraries we found that \emph{OkHttp} and \emph{Retrofit} are used the most.
By far the most used value type in JSON data is \code{STRING}.

We realized that closed-source apps substantially rely on advertisement services, and that they tend to have more complex URL paths consisting of more path segments.
Surprisingly, the secure HTTPS protocol is used in the majority of extracted web APIs from open-source applications, but the opposite is true for closed-source apps.
Obviously, when embedded languages are used along with manual string concatenations, the attack surface for code-injection attacks increases.
Nevertheless, we could identify numerous such cases during the manual examination of the web APIs, \ie embedded SQL and JavaScript content was rather common within web communications.
Even worse, we found many more issues on the server-side: unnecessary disclosure of server configurations, outdated web servers and language interpreters with known security vulnerabilities, leaks of internal error messages, and other sensitive data.
Finally, we also found private APIs without any kind of authentication or authorization mechanisms.

We conclude that a lightweight static code analysis is very helpful in mining web APIs, and that the impact of embedded code in web API requests and the hardening of servers has been deeply underestimated.

\section*{Acknowledgment}
We gratefully acknowledge the financial support of the Swiss National Science Foundation for the project
``Agile Software Assistance'' (SNSF project No.\ 200020-181973, Feb.\ 1, 2019 - April 30, 2022).
We also thank CHOOSE, the Swiss Group for Original and Outside-the-box Software Engineering
of the Swiss Informatics Society, for its financial contribution to the presentation of this paper.

\balance

\bibliographystyle{IEEEtran}
\bibliography{journal,references}

\begin{thebibliography}{10}
\providecommand{\url}[1]{#1}
\csname url@samestyle\endcsname
\providecommand{\newblock}{\relax}
\providecommand{\bibinfo}[2]{#2}
\providecommand{\BIBentrySTDinterwordspacing}{\spaceskip=0pt\relax}
\providecommand{\BIBentryALTinterwordstretchfactor}{4}
\providecommand{\BIBentryALTinterwordspacing}{\spaceskip=\fontdimen2\font plus
\BIBentryALTinterwordstretchfactor\fontdimen3\font minus
  \fontdimen4\font\relax}
\providecommand{\BIBforeignlanguage}[2]{{%
\expandafter\ifx\csname l@#1\endcsname\relax
\typeout{** WARNING: IEEEtran.bst: No hyphenation pattern has been}%
\typeout{** loaded for the language `#1'. Using the pattern for}%
\typeout{** the default language instead.}%
\else
\language=\csname l@#1\endcsname
\fi
#2}}
\providecommand{\BIBdecl}{\relax}
\BIBdecl

\bibitem{Rapoport:2017}
\BIBentryALTinterwordspacing
M.~Rapoport, P.~Suter, E.~Wittern, O.~Lh\'{o}tak, and J.~Dolby, ``Who you gonna
  call?: Analyzing web requests in {Android} applications,'' in
  \emph{Proceedings of the 14th International Conference on Mining Software
  Repositories}, ser. MSR '17.\hskip 1em plus 0.5em minus 0.4em\relax
  Piscataway, NJ, USA: IEEE Press, 2017, pp. 80--90. [Online]. Available:
  \url{https://doi.org/10.1109/MSR.2017.11}
\BIBentrySTDinterwordspacing

\bibitem{Zhou:2015}
Y.~Zhou, L.~Wu, Z.~Wang, and X.~Jiang, ``Harvesting developer credentials in
  {Android} apps,'' in \emph{WISEC}, 2015.

\bibitem{Zuo:2017}
\BIBentryALTinterwordspacing
C.~Zuo and Z.~Lin, ``Smartgen: Exposing server {URL}s of mobile apps with
  selective symbolic execution,'' in \emph{Proceedings of the 26th
  International Conference on World Wide Web}, ser. WWW '17.\hskip 1em plus
  0.5em minus 0.4em\relax Republic and Canton of Geneva, Switzerland:
  International World Wide Web Conferences Steering Committee, 2017, pp.
  867--876. [Online]. Available: \url{https://doi.org/10.1145/3038912.3052609}
\BIBentrySTDinterwordspacing

\bibitem{Rastogi:2016}
V.~Rastogi, R.~Shao, Y.~Chen, X.~Pan, S.~Zou, and R.~Riley, ``Are these ads
  safe: Detecting hidden attacks through the mobile app-web interfaces,'' in
  \emph{NDSS}, 2016.

\bibitem{Mendoza:2018}
A.~Mendoza and G.~Gu, ``Mobile application web {API} reconnaissance:
  Web-to-mobile inconsistencies \& vulnerabilities,'' in \emph{2018 IEEE
  Symposium on Security and Privacy (SP)}.\hskip 1em plus 0.5em minus
  0.4em\relax IEEE, 2018, pp. 756--769.

\bibitem{Ghafari:2017}
M.~Ghafari, P.~Gadient, and O.~Nierstrasz, ``Security smells in {Android},'' in
  \emph{2017 IEEE 17th International Working Conference on Source Code Analysis
  and Manipulation (SCAM)}, Sept 2017, pp. 121--130.

\bibitem{Qiu:2018}
\BIBentryALTinterwordspacing
L.~Qiu, Y.~Wang, and J.~Rubin, ``Analyzing the analyzers: {FlowDroid}/{IccTA},
  {AmanDroid}, and {DroidSafe},'' in \emph{Proceedings of the 27th ACM SIGSOFT
  International Symposium on Software Testing and Analysis}, ser. ISSTA
  2018.\hskip 1em plus 0.5em minus 0.4em\relax ACM, 2018, pp. 176--186.
  [Online]. Available: \url{http://doi.acm.org/10.1145/3213846.3213873}
\BIBentrySTDinterwordspacing

\bibitem{Pauck:2018}
\BIBentryALTinterwordspacing
F.~Pauck, E.~Bodden, and H.~Wehrheim, ``Do {Android} taint analysis tools keep
  their promises?'' in \emph{Proceedings of the 2018 26th ACM Joint Meeting on
  European Software Engineering Conference and Symposium on the Foundations of
  Software Engineering}, ser. ESEC/FSE 2018, 2018, pp. 331--341. [Online].
  Available: \url{http://doi.acm.org/10.1145/3236024.3236029}
\BIBentrySTDinterwordspacing

\bibitem{Corr18a}
\BIBentryALTinterwordspacing
C.~Corrodi, T.~Spring, M.~Ghafari, and O.~Nierstrasz, ``Idea: Benchmarking
  {Android} data leak detection tools,'' in \emph{Engineering Secure Software
  and Systems}, M.~Payer, A.~Rashid, and J.~M. Such, Eds.\hskip 1em plus 0.5em
  minus 0.4em\relax Cham: Springer International Publishing, 2018, pp.
  116--123. [Online]. Available:
  \url{http://scg.unibe.ch/archive/papers/Corr18a.pdf}
\BIBentrySTDinterwordspacing

\bibitem{Winkler:1991}
W.~E. Winkler and Y.~Thibaudeau, \emph{An application of the {Fellegi}-{Sunter}
  model of record linkage to the 1990 {US} decennial census}.\hskip 1em plus
  0.5em minus 0.4em\relax Citeseer, 1991.

\bibitem{Yee:2017}
\BIBentryALTinterwordspacing
G.~O.~M. Yee, ``Model for reducing risks to private or sensitive data,'' in
  \emph{Proceedings of the 9th International Workshop on Modelling in Software
  Engineering}, ser. MISE '17.\hskip 1em plus 0.5em minus 0.4em\relax
  Piscataway, NJ, USA: IEEE Press, 2017, pp. 19--25. [Online]. Available:
  \url{https://doi.org/10.1109/MiSE.2017..6}
\BIBentrySTDinterwordspacing

\bibitem{Gadient:2018}
P.~Gadient, M.~Ghafari, P.~Frischknecht, and O.~Nierstrasz, ``Security code
  smells in {Android} {ICC},'' \emph{Empirical Software Engineering Special
  Issue}, 2018.

\end{thebibliography}

\end{document}